\documentclass[11pt,tightenlines,superscriptaddress,floatfixolumn,english,aps,
notitlepage,nofootinbib]{revtex4-1}

\usepackage{amsmath}
\usepackage{amssymb}
\usepackage{esint}
\usepackage{graphicx}
\usepackage{hyperref}
\usepackage{color}
\usepackage[utf8]{inputenc} 
\usepackage{enumitem}
\usepackage{float}
\usepackage{array}
\usepackage{subfig}
\usepackage{soul}
\usepackage[makeroom]{cancel}

\usepackage[dvipsnames]{xcolor}
\usepackage{mathtools, nccmath}
\usepackage{cleveref}
\usepackage{amsfonts}
\usepackage[normalem]{ulem}

\crefrangeformat{equation}{Eqs. #3(#1)#4--#5(#2)#6}

\newcommand{\mf}{\frac}
\newcommand{\ml}{\left}
\newcommand{\mr}{\right}
\newcommand{\mdfd}{{}_{2}F_2}
\newcommand{\mrdfd}{{{}_{2}\widetilde{F}_2}}
\newcommand{\mfb}{\bar{f}}
\newcommand{\mang}[1]{\langle #1 \rangle}
\newcommand{\mceil}[1]{\left \lceil #1 \right \rceil}

\begin{document}

\title{
Factorial cumulants from short-range correlations and global baryon number conservation \\
}

\author{Micha{\l} Barej}
\email{michal.barej@fis.agh.edu.pl}
\affiliation{AGH University of Science and Technology,
Faculty of Physics and Applied Computer Science,
30-059 Krak\'ow, Poland}

\author{Adam Bzdak}
\email{adam.bzdak@fis.agh.edu.pl}
\affiliation{AGH University of Science and Technology,
Faculty of Physics and Applied Computer Science,
30-059 Krak\'ow, Poland}

\begin{abstract}
We calculate the baryon factorial cumulants 
assuming arbitrary short-range correlations and the global baryon number conservation. The general factorial cumulant generating function is derived. Various relations between factorial cumulants subjected to baryon number conservation and the factorial cumulants without this constraint are presented. We observe that for $n$-th factorial cumulant, the short-range correlations of more than $n$ particles are suppressed with the increasing number of particles. The recently published \cite{Vovchenko:2020tsr} relations between the cumulants in a finite acceptance with global baryon conservation and the grand-canonical susceptibilities are reproduced.
\end{abstract}

\maketitle

\section{Introduction}
Search for the predicted first-order phase transition and the corresponding critical end point between the hadronic matter and quark-gluon plasma is one of the most important challenges in high-energy physics \cite{Stephanov:2004wx,BraunMunzinger:2008tz,Braun-Munzinger:2015hba,Bzdak:2019pkr}. Since fluctuations of such observables as baryon number, electric charge or strangeness number are sensitive to the phase transitions, they are broadly studied both theoretically and experimentally in relativistic heavy-ion collisions \cite{Jeon:2000wg,Asakawa:2000wh,Gazdzicki:2003bb,Gorenstein:2003hk,Stephanov:2004wx,Koch:2005vg,Stephanov:2008qz,Cheng:2008zh,Fu:2009wy,Skokov:2010uh,Stephanov:2011pb,Karsch:2011gg,Schaefer:2011ex,Chen:2011am,Luo:2011rg,Zhou:2012ay,Wang:2012jr,Herold:2016uvv,Luo:2017faz,Szymanski:2019yho,Ratti:2019tvj}. 

Higher-order cumulants, $\kappa_n$, are commonly used to describe such fluctuations \cite{Stephanov:2008qz,Behera:2018wqk,Acharya:2019izy,Skokov:2011rq,BraunMunzinger:2011ta,Bzdak:2012ab,Braun-Munzinger:2016yjz,Adamczyk:2017wsl,Adare:2015aqk}. Nevertheless, the cumulants mix the correlation functions of different orders and also they may be dominated by the trivial average number of particles. On the other hand, the factorial cumulants, $\hat{C}_n$, represent the integrated multiparticle correlation functions \cite{Botet:2002gj,Ling:2015yau,Bzdak:2016sxg,Bzdak:2019pkr} and their applications can be seen, e.g., in Refs. 
\cite{Adamczyk:2013dal,Bzdak:2016sxg,Bzdak:2018uhv,Bzdak:2018axe,HADES:2020wpc,STAR:2021iop,Vovchenko:2021kxx}. However, one should be careful because various effects such as the impact parameter fluctuation or the conservation laws may be reflected in the anomalies of factorial cumulants or cumulants \cite{Skokov:2012ds,Braun-Munzinger:2016yjz,Bzdak:2016jxo,Adamczewski-Musch:2020slf,Kitazawa:2012at,Bzdak:2012an,Braun-Munzinger:2016yjz,Rogly:2018kus,Braun-Munzinger:2019yxj,Acharya:2019izy,Savchuk:2019xfg,Braun-Munzinger:2020jbk,Vovchenko:2021kxx}.

In our previous paper \cite{Barej:2020ymr} we calculated the proton, antiproton and mixed proton-antiproton factorial cumulants  
assuming that the global baryon number conservation is the only source of correlations. We assumed that the acceptance is governed by the binomial distribution, which is correct if there are no other sources of correlations. Recently in Ref. \cite{Vovchenko:2020tsr}, it was argued that applying the binomial acceptance is not correct if, e.g., short-range correlations are present in the system.
Instead of the binomial acceptance, the subensemble acceptance method (SAM) was proposed. 
Using this approach, the relation between cumulants in a finite acceptance with global baryon conservation and the grand-canonical susceptibilities (cumulants), measured, e.g., on the lattice, was derived. The calculation presented in \cite{Vovchenko:2020tsr} assumes that the subvolume, in which cumulants are calculated, is large enough to be close to the thermodynamic limit.

In this paper we use SAM to study the factorial cumulants for one species of particles subjected to short-range correlations and the global baryon number conservation. In particular, we derive the general factorial cumulant generating function and various relations between factorial cumulants subjected to baryon number conservation and cumulants without this constraint. We also observe that for $n$-th factorial cumulant, the short-range correlations of more than $n$ particles are suppressed with the increasing total number of particles. Finally, we reproduce the main results of Ref. \cite{Vovchenko:2020tsr}. 

In the next Section, we present our derivation of the factorial cumulant generating function. In Section \ref{sec:2-particle} we discuss in detail the case of two-particles short-range correlations, providing analytical formulas for the factorial cumulant generating function and the factorial cumulants up to the sixth order. We analyze their dependencies on the correlation strength and acceptance and propose certain approximations. Then, in Section \ref{sec:multi-particle}, we move to multiparticle correlations.  
This is followed by the comparison of the cumulants obtained in our computation with the outcome of Ref. \cite{Vovchenko:2020tsr}. 
Finally, we present our comments and summary.

\section{Factorial cumulant generating function} 
\begin{figure}[H]
\begin{center}
	\includegraphics[width=0.39\textwidth]{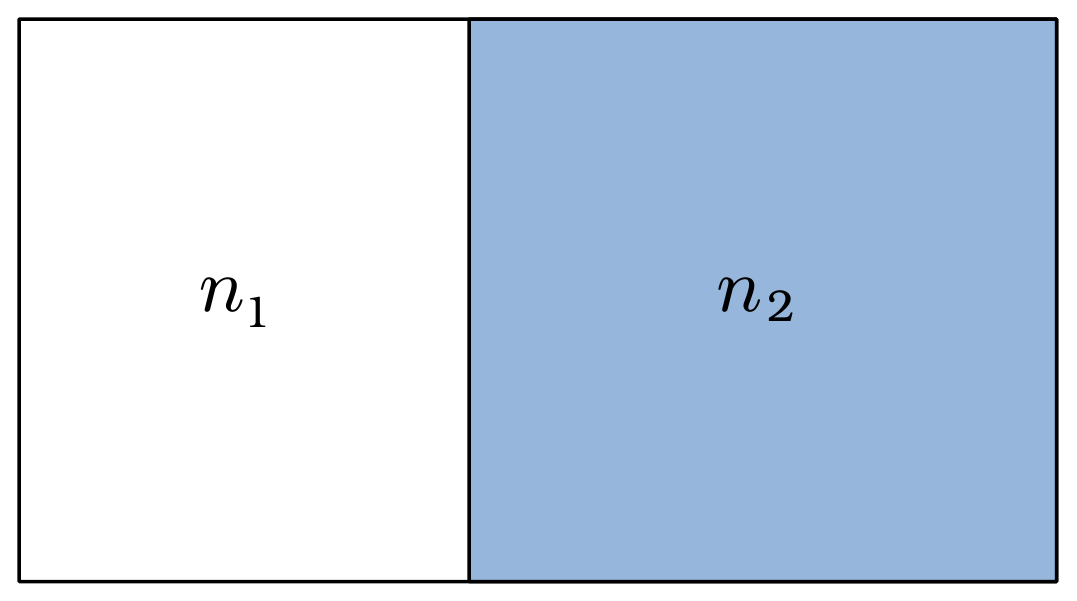}
\caption{The system divided into two subsystems with $n_1$ particles in the first subvolume and $n_2$ particles in the second one. $n_1$ is inside and $n_2$ is outside of our acceptance. } \label{fig:box}
\end{center}
\end{figure}

Consider a system of a fixed volume and some number of baryons of one species only, say protons. We divide it into two subsystems which can exchange particles, see Fig. \ref{fig:box}.
\vspace{-0.2in}
Let $P_1(n_1)$ be the probability that there are $n_1$ baryons in the first subsystem and $P_2(n_2)$ be the probability that there are $n_2$ baryons in the second one. Then, the probability that there are $n_1$ particles in the first part and $n_2$ particles in the second one is given by
\begin{equation}
P(n_1,n_2) = P_1(n_1) P_2(n_2)\,,
\end{equation}
if there are no correlations between the two subsystems. This equation is also approximately true for the case of short-range correlations, that is, if the correlation length is much shorter than the system size. In this paper we assume that this is exactly the case. We note that this is also one of the assumptions of the analysis of Ref. \cite{Vovchenko:2020tsr}.

Now we impose the global baryon number conservation with a fixed total baryon number $B$. In this case 
\begin{equation}
P_B(n_1,n_2) = A\;P_1(n_1) P_2(n_2) \delta_{n_1+n_2,B}\,,
\end{equation}
where $A$ is the normalization constant and $\delta_{n_1+n_2,B}$ is the Kronecker delta responsible for the conservation law, that is $n_1 + n_2 = B$. The probability that there are $n_1$ particles in the first subvolume reads
\begin{equation}
P_{B}(n_1) = \sum_{n_2=0}^{\infty} P_B(n_1,n_2)\,,
\end{equation}
where the subscript $B$ indicates that the quantity is influenced by the conservation law.

The probability generating function corresponding to $P_B(n_1)$ is
\begin{equation}
H_{(1,B)}(z) = \sum_{n_1=0}^{\infty} P_B(n_1)z^{n_1}\,,
\end{equation}
 where here and in the following, the subscript $(1,B)$ indicates that the quantity from the first bin is influenced by the global baryon number conservation. 
Using the integral representation of the Kronecker delta:
\begin{equation}
\delta_{p, r} = \mf{1}{2 \pi i} \oint_{|x| =1} \mf{dx}{x} x^{p-r} \,,
\end{equation}
where $x$ is a complex variable, we obtain:
\begin{equation}
H_{(1,B)}(z) = \mf{A}{2 \pi i} \oint_{|x| =1} \mf{dx}{x^{B+1}} H_1(x z) H_2(x)  \,,
\end{equation}
where
\begin{equation}
H_i(z) = \sum_{n_i =0}^\infty P_i(n_i) z^{n_i}\,, \qquad i=1,2,
\end{equation}
is the probability generating function for the multiplicity distribution $P_i(n_i)$, $i=1,2$, free of the baryon number conservation.

The factorial cumulant generating function is given by (see, e.g., \cite{Bzdak:2019pkr}):
\begin{equation}
G(z) = \ln[H(z)]\,,
\end{equation}
thus,
\begin{equation}
G_{(1,B)}(z) = \ln[H_{(1,B)}(z)] = \ln \ml[ \mf{A}{2 \pi i} \oint_{|x| =1} \mf{dx}{x^{B+1}}\: e^{G_1(x z)} e^{G_2(x)} \mr] \,,
\end{equation}
where $G_1$ and $G_2$ are the factorial cumulant generating functions free of the baryon number conservation.

Using Cauchy's differentiation formula:
\begin{equation}
\oint_{|x| =1} dx \mf{f(x)}{x^{B+1}} = \mf{2 \pi i}{B!} \ml. \mf{d^B f(x)}{dx^B} \mr|_{x=0} \,,
\end{equation}
we obtain
\begin{equation}
G_{(1,B)}(z) = \ln \ml[ \mf{A}{B!} \ml. \mf{d^B}{dx^B} \ml( e^{G_1(x z)} e^{G_2(x)}\mr)\mr|_{x=0} \mr] \,.
\end{equation}

We can express the factorial cumulant generating functions $G_i$ by the series of their factorial cumulants $\hat{C}_k^{(i)}$ ($i=1,2$ denoting the subvolume number). For example\footnote{So that we have $\hat{C}_k^{(2)} = \left. \frac{d^k}{dx^k}G_{2}(x) \right|_{x=1}$.}
\begin{equation} \label{eq:fact-cum-by-series}
G_2(x) = \sum_{k=1}^{\infty} \mf{(x-1)^k}{k!}\hat{C}_k^{(2)}\,.
\end{equation}

This leads to 
\begin{equation}\label{eq:g_b}
G_{(1,B)}(z) = \ln\left[\frac{A}{B!}\left.\frac{d^B}{dx^B}\exp\left( \sum_{k=1}^{\infty} \frac{(xz-1)^k \hat{C}_k^{(1)} + (x-1)^k \hat{C}_k^{(2)}}{k!} \right) \right|_{x=0} \right]\,.
\end{equation}
Note that $\hat{C}_k^{(1)}$, $\hat{C}_k^{(2)}$ are respectively the factorial cumulants 
in the first and the second subsystems for the multiplicity distributions free of the global baryon conservation. As explained earlier, these factorial cumulants are sensitive to the short-range correlations only.

Finally, using the Faà di Bruno's formula (for details see Appendix \ref{appendix:appBruno}) we obtain
\begin{equation} \label{eq:g_b-bell}
   \begin{aligned}
	G_{(1,B)}(z) = \ln\biggl[\frac{A'}{B!} {\rm{Bell}}_B \biggl( & \sum_{k=0}^{\infty}\mf{(-1)^k}{k!} \ml[\hat{C}_{k+1}^{(1)}z + \hat{C}_{k+1}^{(2)}\mr], \\
	& \sum_{k=0}^{\infty}\mf{(-1)^k}{k!} \ml[\hat{C}_{k+2}^{(1)}z^2 + \hat{C}_{k+2}^{(2)}\mr],  \\
	& \ldots, \\
	& \sum_{k=0}^{\infty}\mf{(-1)^k}{k!} \ml[\hat{C}_{k+B}^{(1)} z^B + \hat{C}_{k+B}^{(2)}\mr] \biggr) \biggr] \,,
	\end{aligned}
\end{equation}
where $A'$ is a constant not relevant for further calculations, and 
${\rm{Bell}}_{B}$ is the $B$-th complete exponential Bell polynomial:
\begin{equation} \label{eq:complete-bell}
{\rm{Bell}}_{B}(x_1, x_2, ..., x_B) = \sum_{i=1}^{B} {\rm{Bell}}_{B,i}(x_1, x_2, ..., x_{B-i+1})\,,
\end{equation}
with ${\rm{Bell}}_{B,i}(x_1, x_2, ..., x_{B-i+1})$ being the partial exponential Bell polynomials.
 
The goal of this paper is to relate the factorial cumulants $\hat{C}_k^{(1,B)}$ of $P_{B}(n_1)$
\begin{equation} \label{eq:fact-cum-general}
\hat{C}_k^{(1,B)} = \left.\frac{d^k}{dz^k}G_{(1,B)}(z)\right|_{z=1}\,,
\end{equation}
through the factorial cumulants $\hat{C}_k^{(1)}$, $\hat{C}_k^{(2)}$ of the probability distributions $P_1(n_1)$ and $P_2(n_2)$, respectively.

In this paper we allow for the short-range correlations only (besides the global baryon conservation which results in the long-range correlation) and consequently the factorial cumulants $\hat{C}_k^{(i)}$ (without baryon number conservation) of any order $k$ are proportional to the mean number of particles, see, e.g., \cite{Bzdak:2019pkr}. We have 
\begin{equation} \label{eq:corrs}
\begin{aligned}
\hat{C}_k^{(1)}& = \langle n_1 \rangle \alpha_k = f \langle N \rangle \alpha_k \,, \\
\hat{C}_k^{(2)}& = \langle n_2 \rangle \alpha_k = (1-f) \langle N \rangle \alpha_k\,,
\end{aligned}
\end{equation}
where $\langle N \rangle = \langle n_1 \rangle + \langle n_2 \rangle$ is the mean total number of particles in the system, $f$ is a fraction of particles in the first subvolume, and $\alpha_k$ describes the strength of $k$-particle short-range correlation ($\alpha_1 = 1$). If there are no short-range correlations in the system, then $\alpha_k = 0$ for $k>1$. We note that $\langle n_1 \rangle$ is the mean number of particles of $P_1(n_1)$, that is, the distribution not affected by the global baryon conservation (and analogously for $\langle n_2 \rangle$).

In the following we will usually assume that $\langle N \rangle = B$. 
Introducing the global baryon number conservation further requires that the total number of particles in every event equals $B$. That is why, the average number of baryons with the baryon number conservation included also equals $B$.

\section{Two-particle correlations} \label{sec:2-particle}
Here we consider two-particle short-range correlations only, that is, $\alpha_2 \neq 0$ and $\alpha_k = 0$ for $k \ge 3$. 

\subsection{An analytic approach using the Faà di Bruno's formula and Bell polynomials}
Applying Eqs. (\ref{eq:corrs}) to Eq. (\ref{eq:g_b-bell}) with $\alpha_k = 0$ for $k \ge 3$, we see that only the first two arguments of the $B$-th complete exponential Bell polynomial ${\rm{Bell}}_{B}$ are non-zero, that is 
\begin{equation} \label{eq:g_b-bell-2part}
   \begin{aligned}
	G_{(1,B)}(z) = \ln\biggl[\frac{A'}{B!} {\rm{Bell}}_B \biggl( \sum_{k=0}^{1}\mf{(-1)^k}{k!} \mang{N} \alpha_{k+1} \ml(f \ z + \mfb\, \mr), \;\; \mang{N} \alpha_{2} \ml(f z^2 + \mfb \, \mr), \;\; \underbrace{0, 0, \ldots, 0}_{(B-2) \text{ zeros}} \biggr) \biggr] \,,
	\end{aligned}
\end{equation}
where $\mfb = 1-f$.

Using Eq. (\ref{eq:complete-bell}) we can rewrite it as follows\footnote{Here the last $(B-i-1)$ arguments of ${\rm{Bell}}_{B,i}$ are zeros since ${\rm{Bell}}_{B,i}$ has $(B - i + 1)$ arguments. In particular, ${\rm{Bell}}_{B,B-1}$ has two arguments and no zeros, ${\rm{Bell}}_{B,B-2}$ has three arguments including one being zero, etc. For clarity we separate ${\rm{Bell}}_{B,B}$ because it has one argument only.} 
\begin{equation} 
   \begin{aligned}
	G_{(1,B)}(z) = \ln\biggl\{\frac{A'}{B!} &\biggl[{\rm{Bell}}_{B,B}\ml( (f z + \mfb) \mang{N}(1 -\alpha_2) \mr) \\
	 &+ \sum_{i=1}^{B-1} {\rm{Bell}}_{B,i} \biggl( (f z + \mfb\,) \mang{N} (1 - \alpha_2), \;\;
	(f z^2 + \mfb \,) \mang{N} \alpha_2, \;  
	\underbrace{0, 0, \ldots, 0}_{(B-i-1) \text{ zeros}} \biggr) \biggr] \biggr\}  \,.
	\end{aligned}
\end{equation}

In the next step we apply the definition of the partial exponential Bell polynomials:
\begin{equation} \label{eq:g_b-bell-partial-def}
{\rm{Bell}}_{B,i} (x_1, x_2, ..., x_{B-i+1}) = \sum \mf{B!}{j_1! j_2! \cdots j_{B-i+1}!} \ml( \mf{x_1}{1!} \mr)^{j_1} \ml( \mf{x_2}{2!} \mr)^{j_2} \cdots \ml( \mf{x_{B-i+1}}{(B-i+1)!} \mr)^{j_{B-i+1}}\,,
\end{equation}
where the sum is over the non-negative integer $j_1, j_2, ..., j_{B-i+1}$ such that 
\begin{equation} \label{eq:j-sums}
\begin{aligned}
&j_1 + j_2 + ... + j_{B-i+1} = i\,,\\
&j_1 + 2 j_2 + 3 j_3 + ... + (B-i+1) j_{B-i+1} = B\,.
\end{aligned}
\end{equation}
However in our case: $x_3 = x_4 = ... = x_{B-i+1} = 0$, so we have non-zero terms in Eq. (\ref{eq:g_b-bell-partial-def}) if and only if $j_3 = j_4 = ... = j_{B-i+1} = 0$ (because $0^r \neq 0$ for $r=0$ only). In this case, the constraints (\ref{eq:j-sums}) lead to $j_1 = 2i-B$, $j_2 = B - i$. 
Since both $j_1$ and $j_2$ have to be greater than or equal to 0, meaning $B/2 \le i \le B$, 
we obtain\footnote{This result naturally includes ${\rm{Bell}}_{B,B}\ml( (f z + \mfb) \mang{N}(1 -\alpha_2) \mr) = \ml( (f z + \mfb) \mang{N}(1 -\alpha_2) \mr)^B$.}
\begin{equation} 
   \begin{aligned}
	G_{(1,B)}(z) = \ln\biggl[\frac{A'}{B!} \sum_{i=B_0}^{B} \mf{B!}{(2i-B)! (B-i)!} \;  \ml[(f z + \mfb\,) \mang{N}(1 - \alpha_2) \mr]^{2i-B}
	 \ml[ \tfrac{1}{2} (f z^2 + \mfb\,) \mang{N} \alpha_2 \mr]^{B-i}\biggr]\,,
	\end{aligned}
\end{equation}
where 
\begin{equation}
B_0 = \mceil{\mf{B}{2}} \equiv \begin{cases}
\mf{B}{2}, & \text{for } B \text{ even}\\
\mf{B+1}{2}, & \text{for } B \text{ odd}
\end{cases}\,.
\end{equation}
Taking $\langle N \rangle = B$ 
we can rewrite $G_{(1,B)}(z)$ as 
\begin{equation} 
   \begin{aligned} \label{eq:gfun-2f2}
	G_{(1,B)}(z) = \ln\biggl[ & A'  \mf{[B(1-\alpha_2) (f z + \mfb)]^{2B_0-B} \ml[\mf{1}{2}B \alpha_2 (f z^2 + \mfb) \mr]^{B-B_0} }{ (B-B_0)! } \\ 
	& \times {}_{2}F_2\ml(1, B_0-B; B_0 - \mf{B}{2}+ \mf{1}{2}, B_0 -\mf{B}{2} + 1; -\mf{B(1-\alpha_2)^2 (f z + \mfb)^2}{2\alpha_2 (f z^2 + \mfb)} 
	\mr) \biggr]\,,
	\end{aligned}
\end{equation}
where ${}_{2}F_2(...)$ is the generalized hypergeometric function defined as
\begin{equation}  \label{eq:pfq}
{}_{2}F_2(a_1, a_2; b_1, b_2; z) = \sum_{n=0}^{\infty}\mf{(a_1)^{(n)}(a_2)^{(n)}}{(b_1)^{(n)}(b_2)^{(n)}} \mf{z^n}{n!}
\end{equation}
with
\begin{equation} 
(x)^{(n)} =  \begin{cases}
			\prod_{k=0}^{n-1} (x + k), & \text{if $n=1,2,3,...$ }\\
            1, & \text{if $n=0$}
		 \end{cases}
\end{equation}
being the rising factorial (Pochhammer symbol).

Note that for even $B$, the fourth argument of ${}_{2}F_2$ becomes 1 whereas for odd $B$ the third argument becomes 1. Therefore, ${}_{2}F_2$ is in either case reduced to ${}_{1}F_1$ (the confluent hypergeometric function). Moreover, $(B_0 - B)$, the second argument of ${}_{2}F_2$ is a negative integer, so $(B_0 - B)^{(n)} = 0$ starting from $n = B - B_0 + 1$. Therefore, the sum in ${}_{2}F_2$ is in fact finite from $n=0$ to $B - B_0$.\footnote{It is straightforward to show that for the special case of $\alpha_2 = 0$ (no short-range correlations, making the global baryon number conservation the only source of correlations), the factorial cumulant generating function, Eq. (\ref{eq:gfun-2f2}), 
becomes $G_{(1,B)}(z) = \widetilde{C} + B \ln(f z + \mfb)$, where $\widetilde{C} = \ln\ml(\mf{A}{B!} B^B \mr) - B$. Obviously the same result is obtained assuming $\alpha_k = 0$ for $k \geq 2$ already in Eq. (\ref{eq:g_b}). The resulting factorial cumulants are $\hat{C}_n^{(1,B)} = (-1)^{n-1}(n-1)! f^n B$, consistent with the approach presented in Ref. \cite{Barej:2020ymr} but applied to one kind of particles.}

The factorial cumulants obtained from Eq. (\ref{eq:gfun-2f2}), see Eq. (\ref{eq:fact-cum-general}), read: 
\begin{flalign}\label{eq:c1-bell}
	\hat{C}_1^{(1,B)} &= fB  \,,&&
\end{flalign}
\begin{flalign}\label{eq:c2-bell}
\hat{C}_2^{(1,B)} &= - f^2 B + 2f \beta - \mf{f \beta \gamma}{\alpha_2} R_1  \,,&&
\end{flalign} 
\begin{flalign}\label{eq:c3-bell}
	\hat{C}_3^{(1,B)} &= 2 f^3 B -12f^2 \beta + 6\mf{f^2 \beta \gamma}{\alpha_2} R_1  \,,&&
\end{flalign} 
\begin{flalign}\label{eq:c4-bell}
	\hat{C}_4^{(1,B)} &= - 3!f^4 B -12f^2(1-7f)\beta  -3 \mf{f^2 \beta \gamma}{\alpha_2^2} \ml[ \beta \gamma R_1^2 - 4(1-4f)\alpha_2 R_1 - 2 \mfb \gamma (B-B_0-1) R_2 \mr] &&
\end{flalign} 
\begin{flalign}\label{eq:c5-bell}
	\hat{C}_5^{(1,B)} &= 4!f^5 B + 240f^3(1-3f)\beta + 60 \mf{f^3 \beta \gamma}{\alpha_2^2} \ml[ \beta \gamma R_1^2 - 4(1-2f)\alpha_2 R_1 - 2 \mfb \gamma (B-B_0-1) R_2 \mr] &&
\end{flalign} 
\begin{flalign}\label{eq:c6-bell}
	\hat{C}_6^{(1,B)} &= -5!f^6 B + 60\mf{f^3 \beta}{\alpha_2} \ml[- \mfb^2 \gamma   + 4(31f^2-17f+1)\alpha_2 \mr] && \\ \nonumber
	&-15 \mf{f^3 \beta \gamma}{\alpha_2^3} \biggl[ 2\beta^2 \gamma^2 R_1^3 -6 \mfb \beta \gamma^2 (B-B_0-1)  R_1 R_2 -12 \beta \gamma(1-6f)\alpha_2 R_1^2 && \\ \nonumber
	&\phantom{-15 \mf{f^3 \gamma}{\alpha_2^3} \biggl[} + 2\alpha_2^2 \ml[191f^2 - 142f +11 - B_0(3+2B_0)\mfb^2 \mr] R_1&& \\ \nonumber
	&\phantom{-15 \mf{f^3 \gamma}{\alpha_2^3} \biggl[} + \alpha_2 \mfb^2 B \ml[(2\alpha_2^2 -5\alpha_2 + 2)B -2(\alpha_2^2-4\alpha_2+1)B_0 -(4\alpha_2^2-11\alpha_2+4) \mr] R_1 &&  \\ \nonumber
	&\phantom{-15 \mf{f^3 \gamma}{\alpha_2^3} \biggl[} + 2 \mfb \gamma (B-B_0-1) \ml[(7-4B_0 \mfb-67f)\alpha_2 - B \mfb (1-4\alpha_2 + \alpha_2^2)\mr] R_2 \biggr]\,, &&
\end{flalign}
where the commonly appearing terms are defined as follows:
\begin{equation}
	\beta = (B-B_0) \mfb \,,
\end{equation}
\begin{equation}
	\gamma = B(1-\alpha_2)^2 \,,
\end{equation}
\begin{equation}
	R_n = \mf{\mrdfd\ml(1 + n, B_0 - B+ n; B_0 - \mf{B}{2} + \mf{1}{2} + n, B_0-\mf{B}{2} + 1 + n;- \mf{B(1-\alpha_2)^2}{2\alpha_2}\mr)}{\mrdfd\ml(1, B_0 - B; B_0 - \mf{B}{2} + \mf{1}{2} , B_0-\mf{B}{2} + 1; - \mf{B(1-\alpha_2)^2}{2\alpha_2}\mr)} \,,
\end{equation}
whith
\begin{equation}
\mrdfd(a_1, a_2; b_1, b_2, z) = \mf{\mdfd(a_1, a_2; b_1, b_2, z)}{\Gamma(b_1)\Gamma(b_2)} \,
\end{equation}
being the regularized generalized hypergeometric function.

\subsection{An approximate approach using general Leibnitz formula}
In the case of two-particle correlations only, Eq. (\ref{eq:g_b}) reads: 
\begin{equation}\label{eq:g_b-for-2}
\begin{aligned}
G_{(1,B)}(z) = \ln\biggl[\frac{A}{B!} \biggl.\frac{d^B}{dx^B}& \biggl[ \exp\biggl((xz-1)f \langle N \rangle + (x-1)\mfb\langle N \rangle \biggr) \times \\ & \,\, \exp\biggl(\frac{1}{2}(xz-1)^2 f\langle N \rangle \alpha_2 + \frac{1}{2}(x-1)^2 \mfb\langle N \rangle \alpha_2\biggr)  \biggr]\biggr|_{x=0}\biggr]\,,
\end{aligned}
\end{equation}
where $\alpha_2$ is defined in Eq. (\ref{eq:corrs}). 

Assuming $\alpha_2$ to be small, we can expand the second exponent in Eq. (\ref{eq:g_b-for-2}) into a series. Then, we calculate the $B$-th derivative of a product of two functions, which is given by the general Leibnitz formula:
\begin{equation}\label{eq:general-leibnitz}
\frac{d^B(u\,v)}{dx^B} = \sum_{k=0}^B \binom{B}{k} \frac{d^{B-k}u}{dx^{B-k}} \frac{d^k v}{dx^k} \,.
\end{equation}
In our case: 
\begin{equation} \label{eq:u-fun}
u(x)=\exp\ml((xz-1)f \langle N \rangle + (x-1)\mfb\langle N \rangle \mr)\,,
\end{equation}
\begin{equation}
\begin{aligned} \label{eq:v-fun}
v(x)&=\exp\ml(\tfrac{1}{2}(xz-1)^2 f\langle N \rangle \alpha_2 + \tfrac{1}{2}(x-1)^2 \mfb\langle N \rangle \alpha_2\mr) \\
	&\approx \sum_{m=0}^{m_{max}} \mf{1}{m!} \ml( \tfrac{1}{2}(xz-1)^2 f\langle N \rangle \alpha_2 + \tfrac{1}{2}(x-1)^2 \mfb\langle N \rangle \alpha_2 \mr)^m \,,
\end{aligned}
\end{equation}
\begin{equation}
\frac{d^k u}{dx^k}=\left(\mfb\langle N \rangle + f\,z\langle N \rangle  \right)^k \exp\ml((xz-1)f \langle N \rangle + (x-1)\mfb\langle N \rangle \mr)\,,
\end{equation}
\begin{equation}
\frac{d^k v}{dx^k}=0 \qquad \text{for } k>2m_{max}\,.
\end{equation}
Taking $\langle N \rangle = B$ we obtain:
\begin{equation} 
   \begin{aligned} \label{eq:2-particle-gfun}
G_{(1,B)}(z) & =\ln(A)-\ln(B!)-B+B\ln(BY_{1})\\
 & +\ln\biggl[1\\
 & \phantom{+\ln}+\tfrac{1}{2}B\biggl(\frac{B-1}{B}X_2-1\biggr)\;\alpha_{2}\\
 & \phantom{+\ln}+\tfrac{1}{4!!}B^{2}\biggl(\frac{B!}{(B-4)!}\frac{X_2^{2}}{B^{4}}-4\frac{B!}{(B-3)!}\frac{X_2}{B^{3}}+2\frac{B!}{(B-2)!}\frac{X_2+2}{B^{2}}-3\biggr)\;\alpha_{2}^{2}\\
 & \phantom{+\ln}+\tfrac{1}{6!!}B^{3}\biggl(\frac{B!}{(B-6)!}\frac{X_2^{3}}{B^{6}}-6\frac{B!}{(B-5)!}\frac{X_2^{2}}{B^{5}}+3\frac{B!}{(B-4)!}\frac{X_2(X_2+4)}{B^{4}}\\
 & \phantom{+\ln+\tfrac{1}{48}\langle N \rangle^{3}}-4\frac{B!}{(B-3)!}\frac{3X_2+2}{B^{3}}+3\frac{B!}{(B-2)!}\frac{X_2+4}{B^{2}}-5\biggr)\;\alpha_{2}^{3}\\
 & \phantom{+\ln}+\tfrac{1}{8!!}B^{4}\biggl(\frac{B!}{(B-8)!}\frac{X_2^{4}}{B^{8}}-8\frac{B!}{(B-7)!}\frac{X_2^{3}}{B^{7}}+4\frac{B!}{(B-6)!}\frac{X_2^{2}(X_2+6)}{B^{6}}-8\frac{B!}{(B-5)!}\frac{X_2(3X_2+4)}{B^{5}}\\
 & \phantom{+\ln+\tfrac{1}{48}\langle N \rangle^{3}}+2\frac{B!}{(B-4)!}\frac{3X_2(X_2+8)+8}{B^{4}}-8\frac{B!}{(B-3)!}\frac{3X_2+4}{B^{3}}+4\frac{B!}{(B-2)!}\frac{X_2+6}{B^{2}}-7\biggr)\;\alpha_{2}^{4}\\
 & \phantom{+\ln}+\ldots\biggr]\,,
   \end{aligned}
\end{equation}
where $X_k = Y_k/(Y_1)^k$ and $Y_k=f z^k + \mfb$ (in this formula only $X_2$ and $Y_1$ appear).

Note that for large $B$ and very small $\alpha_2$ we obtain a simple formula:
\begin{equation} \label{eq:simple-2-particle}
G_{(1,B)}(z) \approx \ln(A)-\ln(B!)-B+B\ln(BY_{1}) +\tfrac{1}{2} (X_2-1) B\; \alpha_2 \,.
\end{equation}

Using Eq. (\ref{eq:fact-cum-general}) we calculate $\hat{C}_n^{(1,B)}$ and again expand into the Taylor series in $\alpha_2$. We obtain:
	\begin{flalign}\label{eq:c1}
	\hat{C}_1^{(1,B)} &= fB  \,,&&
	\end{flalign}
	\begin{flalign}\label{eq:c2}
	\hat{C}_2^{(1,B)} &= -f^2B + \mfb f (B-1)\;\alpha_2 + 3\mfb f\frac{B-1}{B}\;\alpha_2^2 -5\mfb f\frac{(B-1)(B-3)}{B^2}\;\alpha_2^3 \\ \nonumber
	&+ \mfb f\frac{(B-1)(7B^2-65B+105)}{B^3}\;\alpha_2^4 + \ldots  \,,&&
	\end{flalign} 
	\begin{flalign}\label{eq:c3}
	\hat{C}_3^{(1,B)} &= 2f^3B - 6\mfb f^2 (B-1)\;\alpha_2 - 18\mfb f^2\frac{B-1}{B}\;\alpha_2^2 + 30\mfb f^2\frac{(B-1)(B-3)}{B^2}\;\alpha_2^3 \\ \nonumber
	&- 6\mfb f^2\frac{(B-1)(7B^2-65B+105)}{B^3}\;\alpha_2^4 + \ldots  \,,&&
	\end{flalign}
	\begin{flalign}\label{eq:c4}
	\hat{C}_4^{(1,B)} &= -3!f^4B + 36\mfb f^3 (B-1)\;\alpha_2 - 6\mfb f^2(B-1)\frac{2\mfb B-15f - 3}{B}\;\alpha_2^2 \\ \nonumber &
	+ 12\mfb f^2(B-1)\frac{\mfb B^2-B(4f+11)+15(1+2f)}{B^2}\;\alpha_2^3 \\ \nonumber &
	- 6\mfb f^2(B-1)\frac{2\mfb B^3-21B^2(3-f)+15B(19+7f)-315(1+f)}{B^3}\;\alpha_2^4+ \ldots  \,,&&
	\end{flalign}
	\begin{flalign}
	\hat{C}_5^{(1,B)} &= 4!f^5B - 240\mfb f^4(B-1)\;\alpha_2 +120\mfb f^3(B-1)\frac{2\mfb B-3f - 3}{B}\;\alpha_2^2  \\ \nonumber &
	-240 \mfb f^3(B-1)\frac{\mfb B^2-B(11-6f)+15}{B^2}\;\alpha_2^3 \\ \nonumber &
	+ 120\mfb f^3(B-1)\frac{2\mfb B^3-7B^2(9-7f)+5B(57-31f)-105(3-f)}{B^3}\;\alpha_2^4 + \ldots  \,,&&
	\end{flalign}
	\begin{flalign}\label{eq:c6}
	\hat{C}_6^{(1,B)} &= -5!f^6B +1800\mfb f^5(B-1)\;\alpha_2 -1800\mfb f^4(B-1)\frac{2\mfb B - 3}{B}\;\alpha_2^2\\ \nonumber &
	-120 \mfb f^3(B-1)\frac{-5\mfb(5f+1)B^2 +(-238f^2 + 296f + 17)B +15(14f^2 - 28f - 1)}{B^2}\;\alpha_2^3 \\ \nonumber &
	-360\mfb f^3(B-1)\\ \nonumber& 
	\times \frac{2\mfb(2+3f) B^3 +5B^2(47f^2-45f-9) -25B(39f^2 -47f -5)+105(9f^2-13f-1)}{B^3}\;\alpha_2^4 + \ldots  \,.&&
	\end{flalign}

Here we expand up to $\alpha_{2}^{4}$ but higher orders can be easily obtained.\footnote{Obviously, in order to calculate more terms, one needs to take larger $m_{max}$ in Eq. (\ref{eq:v-fun}).}

We note that using Eq. (\ref{eq:simple-2-particle}) we simply obtain the first two terms in each $\hat{C}_k^{(1,B)}$ of \Crefrange{eq:c1}{eq:c6} (with $B-1 \approx B$).

Having factorial cumulants, one can easily calculate cumulants $\kappa_n$.\footnote{Relations between cumulants and factorial cumulants read: $\kappa_2 = \langle n \rangle + \hat{C}_2$, $\kappa_3 = \langle n \rangle + 3\hat{C}_2 + \hat{C}_3$, $\kappa_4 = \langle n \rangle + 7\hat{C}_2 + 6\hat{C}_3 + \hat{C}_4$, $\kappa_5 = \langle n \rangle + 15\hat{C}_2 + 25\hat{C}_3 + 10\hat{C}_4 + \hat{C}_5$, $\kappa_6 = \langle n \rangle + 31\hat{C}_2 + 90\hat{C}_3 + 65\hat{C}_4 + 15\hat{C}_5 + \hat{C}_6$, where mean $\langle n \rangle = \hat{C}_1 = \kappa_1$. Details can be found, e.g., in Appendix A of Ref. \cite{Bzdak:2019pkr}.} For example:
	\begin{flalign}\label{eq:k1}
	\kappa_1^{(1,B)} &= fB  \,,&&
	\end{flalign}
	\begin{flalign}
	\kappa_{2}^{(1,B)} & =\mfb f\biggl[B+(B-1)\;\alpha_{2}+3\frac{B-1}{B}\;\alpha_{2}^{2}-5\frac{(B-1)(B-3)}{B^{2}}\;\alpha_{2}^{3}\\ \nonumber
	&\phantom{=f\mfb} \: +\frac{(B-1)(7B^{2}-65B+105)}{B^{3}}\;\alpha_{2}^{4}+\ldots\biggr]\,,&&
	\end{flalign}
	\begin{flalign}
	\kappa_3^{(1,B)} &= \mfb f(1-2f)\biggl[B+3(B-1)\;\alpha_{2}+9\frac{B-1}{B}\;\alpha_{2}^{2}-15\frac{(B-1)(B-3)}{B^{2}}\;\alpha_{2}^{3}\\ \nonumber
	&\phantom{=\mfb f(1-2f)} \:+3\frac{(B-1)(7B^{2}-65B+105)}{B^{3}}\;\alpha_{2}^{4}+\ldots\biggr]  \,,&&
	\end{flalign}
	\begin{flalign}\label{eq:k4}
	\kappa_4^{(1,B)} &= \mfb f\biggl[B(1-6\mfb f)+(B-1)[7-36\mfb f]\alpha_{2}+3\frac{B-1}{B}[7-2\mfb f(2B+15)]\alpha_{2}^{2}\\ \nonumber
	&\phantom{=\mfb f} \:-\frac{B-1}{B^{2}}[35(B-3)-12\mfb f(B^{2}+4B-30)]\alpha_{2}^{3}\\ \nonumber
	&\phantom{=\mfb f} \:+\frac{B-1}{B^{3}}[7(7B^{2}-65B+105) -6\mfb f(2B^{3}-21(B^{2}+5B-15))]\alpha_{2}^{4} +\ldots\biggr] \,.&&
	\end{flalign}

\subsection{Examples}
\begin{figure}[H]
\begin{center}
	\includegraphics[width=0.49\textwidth]{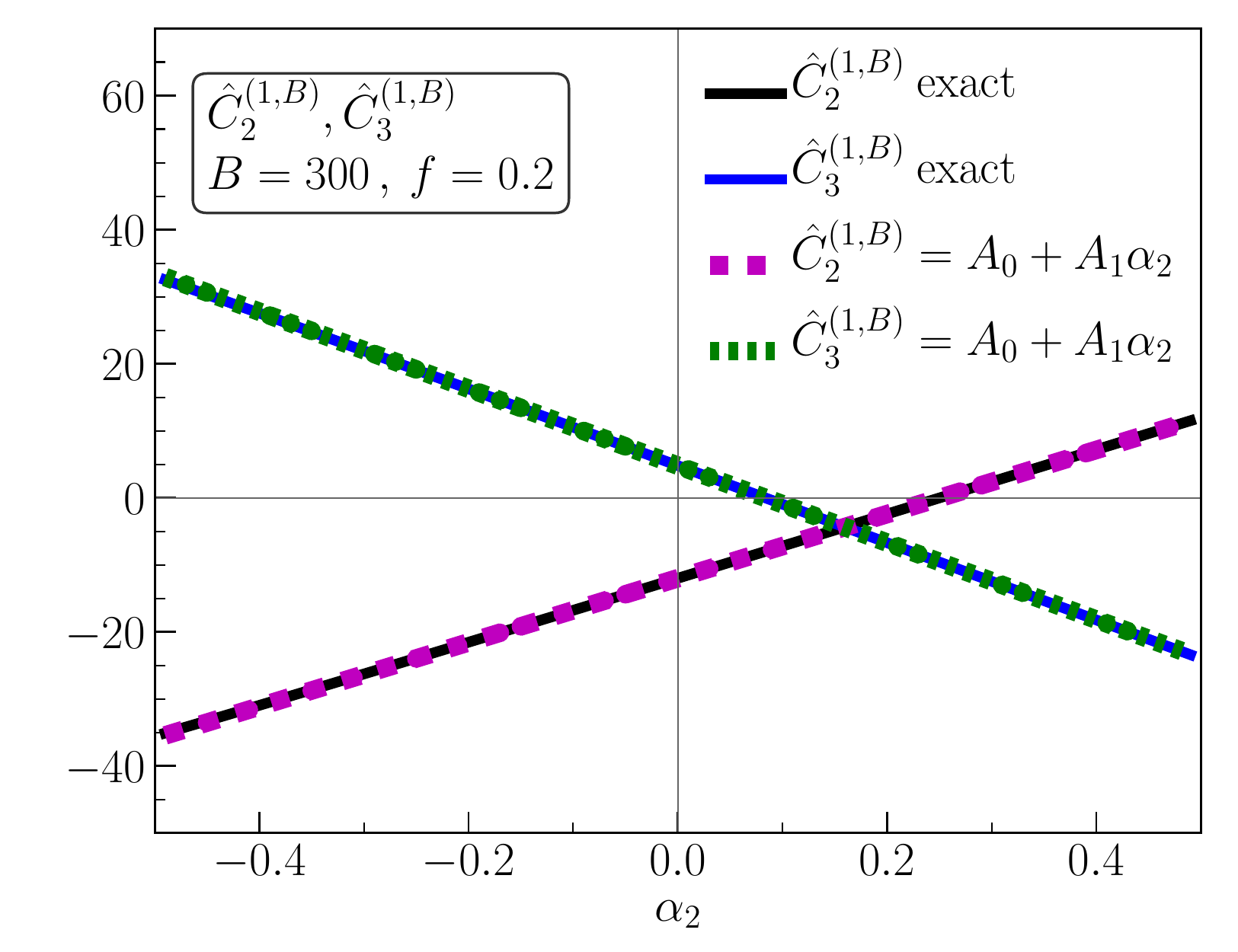}
	\includegraphics[width=0.49\textwidth]{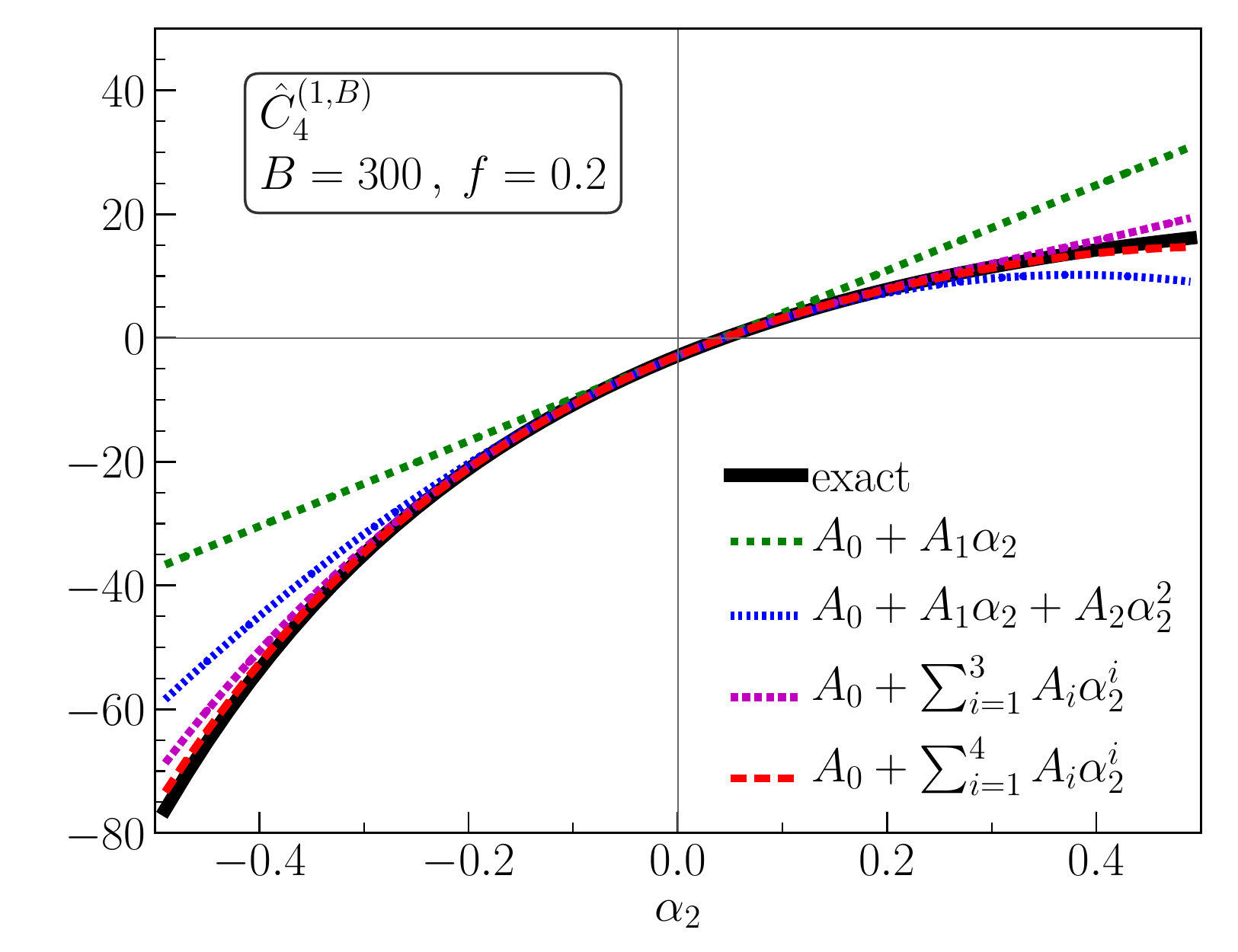}
	\includegraphics[width=0.49\textwidth]{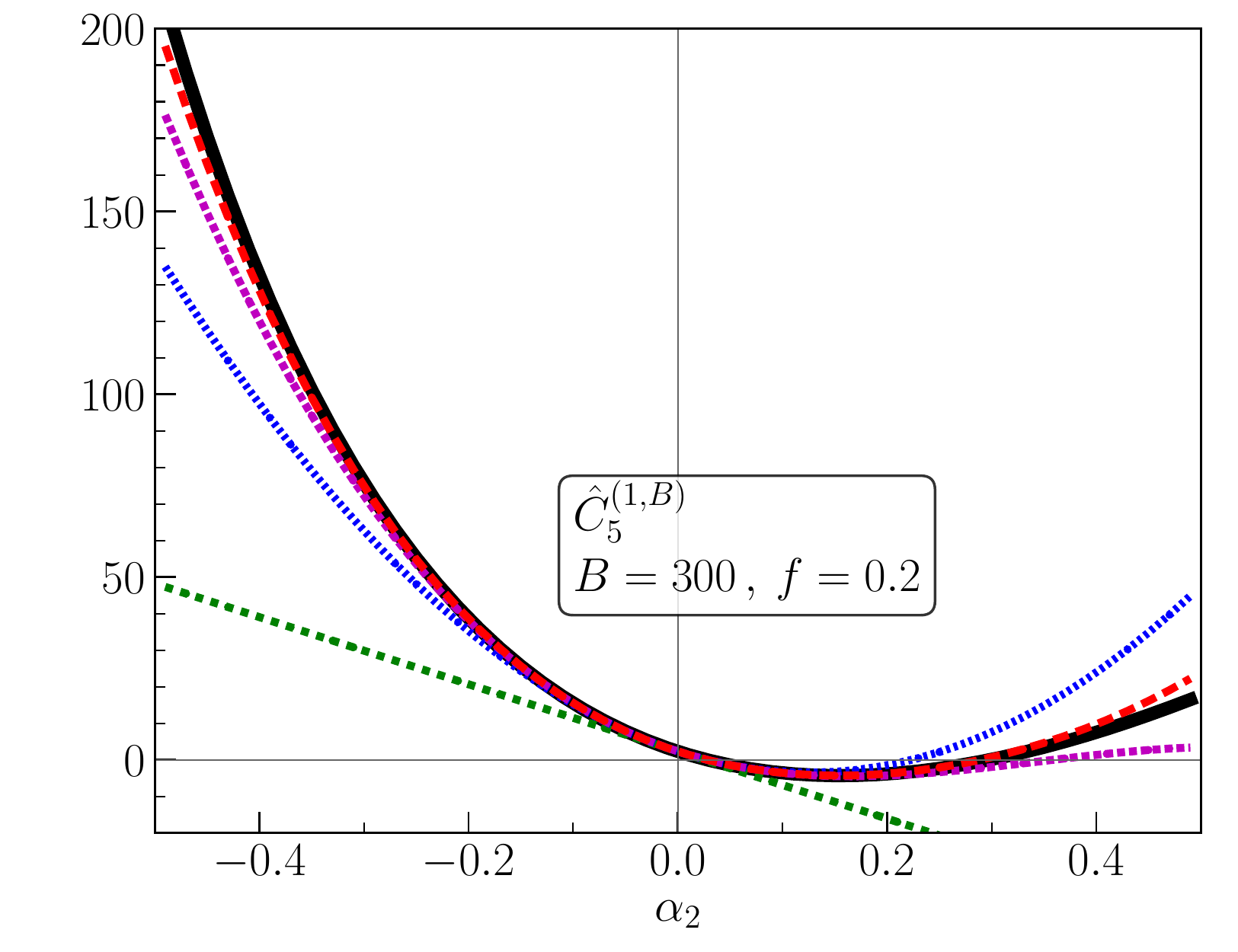}
	\includegraphics[width=0.49\textwidth]{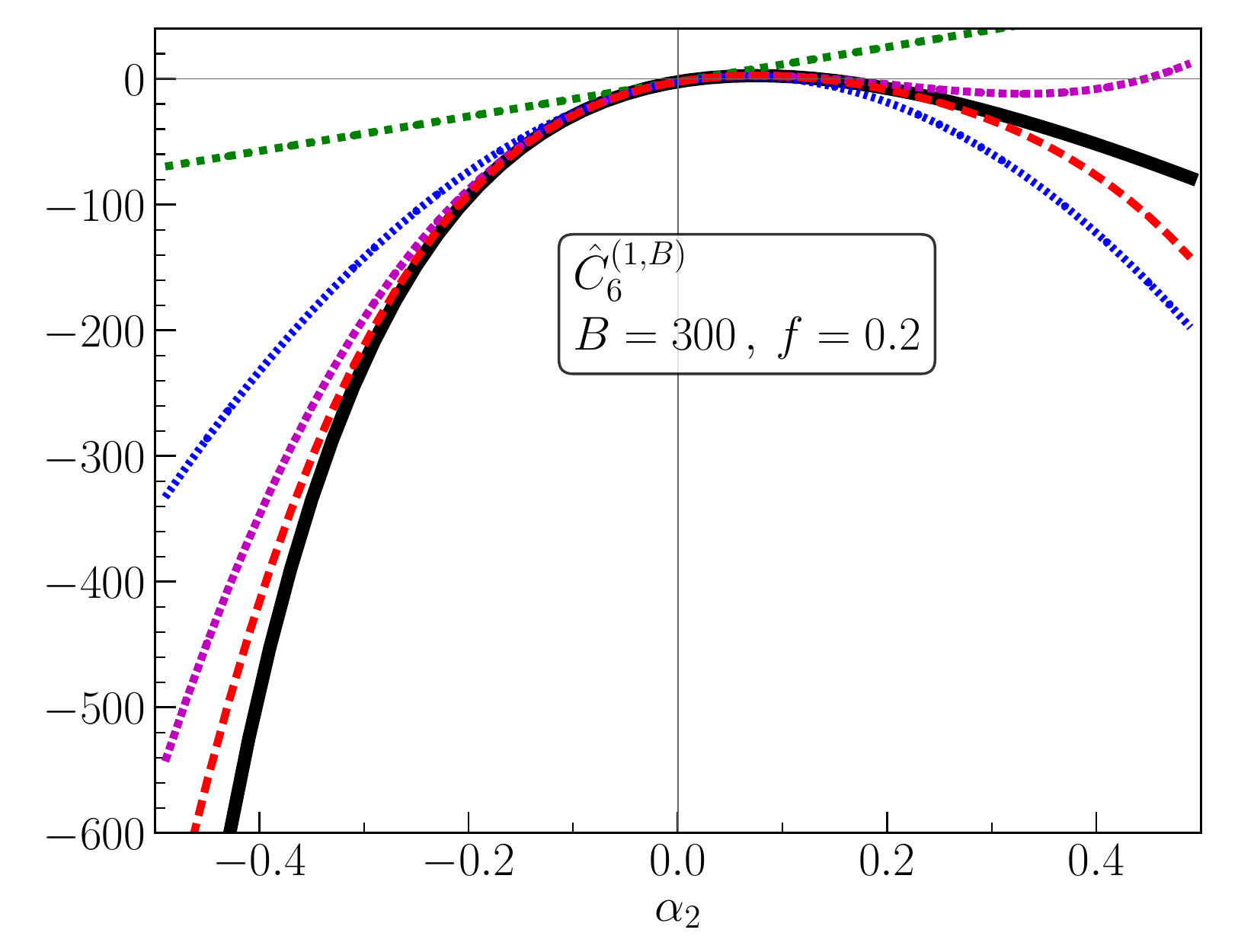}
\caption{Factorial cumulants $\hat{C}_n^{(1,B)}$ ($n=2,3,...,6$, assuming $B=300$, $f=0.2$, and $\alpha_k = 0$ for $k \geq 3$) as a function of the two-particle correlation strength, $\alpha_2$. The ``exact'' line in each plot represents the analytic result, Eqs. (\ref{eq:c1-bell})-(\ref{eq:c6-bell}). Other lines represent approximate power series expansions, Eqs. (\ref{eq:c1})-(\ref{eq:c6}). For $\hat{C}_2^{(1,B)}$ and $\hat{C}_3^{(1,B)}$ only linear terms are presented because higher powers of $\alpha_2$ give practically identical results.} \label{fig:c-vs-alpha}
\end{center}
\end{figure}

To illustrate our results, in Fig. \ref{fig:c-vs-alpha} we plot the factorial cumulants for $B=300$ and $f=0.2$ as a function of $\alpha_2$. We verified that the analytic results obtained using the Faà di Bruno's formula and Bell polynomials, Eqs. (\ref{eq:c1-bell})-(\ref{eq:c6-bell}), are equivalent to the exact computation obtained with differentiating Eq. (\ref{eq:g_b-for-2}) $B=300$ times. We compared them with approximate results obtained using general Leibnitz formula, Eqs. (\ref{eq:c1})-(\ref{eq:c6}).  
As seen in Fig. \ref{fig:c-vs-alpha}, $\hat{C}_2^{(1,B)}$ and $\hat{C}_3^{(1,B)}$ are very well approximated already by a linear expansion in $\alpha_2$. This is not surprising. It is clear from Eqs. (\ref{eq:c2}) and (\ref{eq:c3}) that higher powers of $\alpha_2$ are suppressed for large $B$.
Quartic expansions of $\hat{C}_4^{(1,B)}$ and $\hat{C}_5^{(1,B)}$ are in good agreement with exact results in the whole investigated range $-0.5 \le \alpha_2 \le 0.5$, whereas quartic expansion of $\hat{C}_6^{(1,B)}$ works in a narrower range of $-0.3 \lesssim \alpha_2 \lesssim 0.3$, which is acceptable since $\alpha_2$ was assumed to be rather small.

\begin{figure}[H]
\begin{center}
	\includegraphics[width=0.49\textwidth]{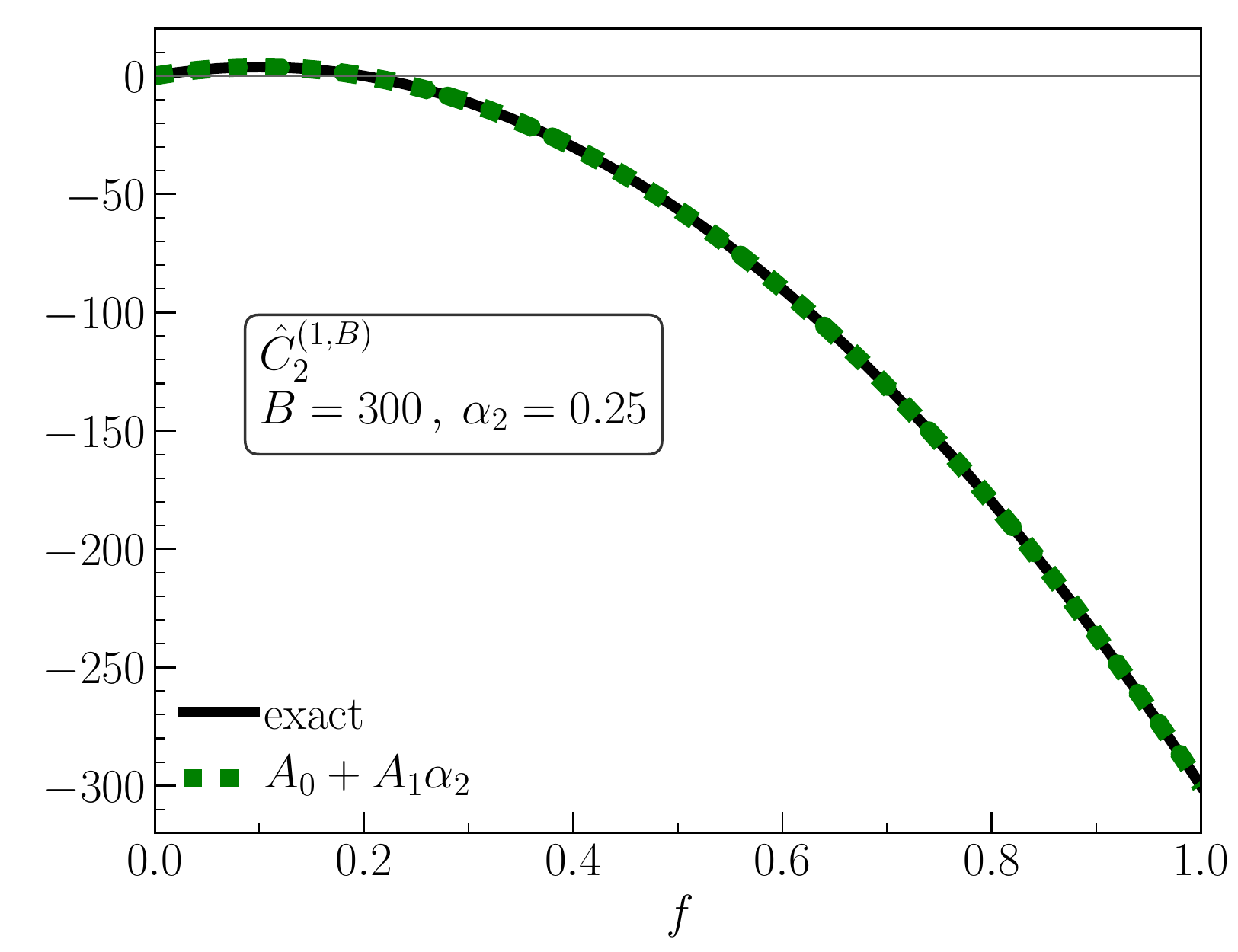}
	\includegraphics[width=0.49\textwidth]{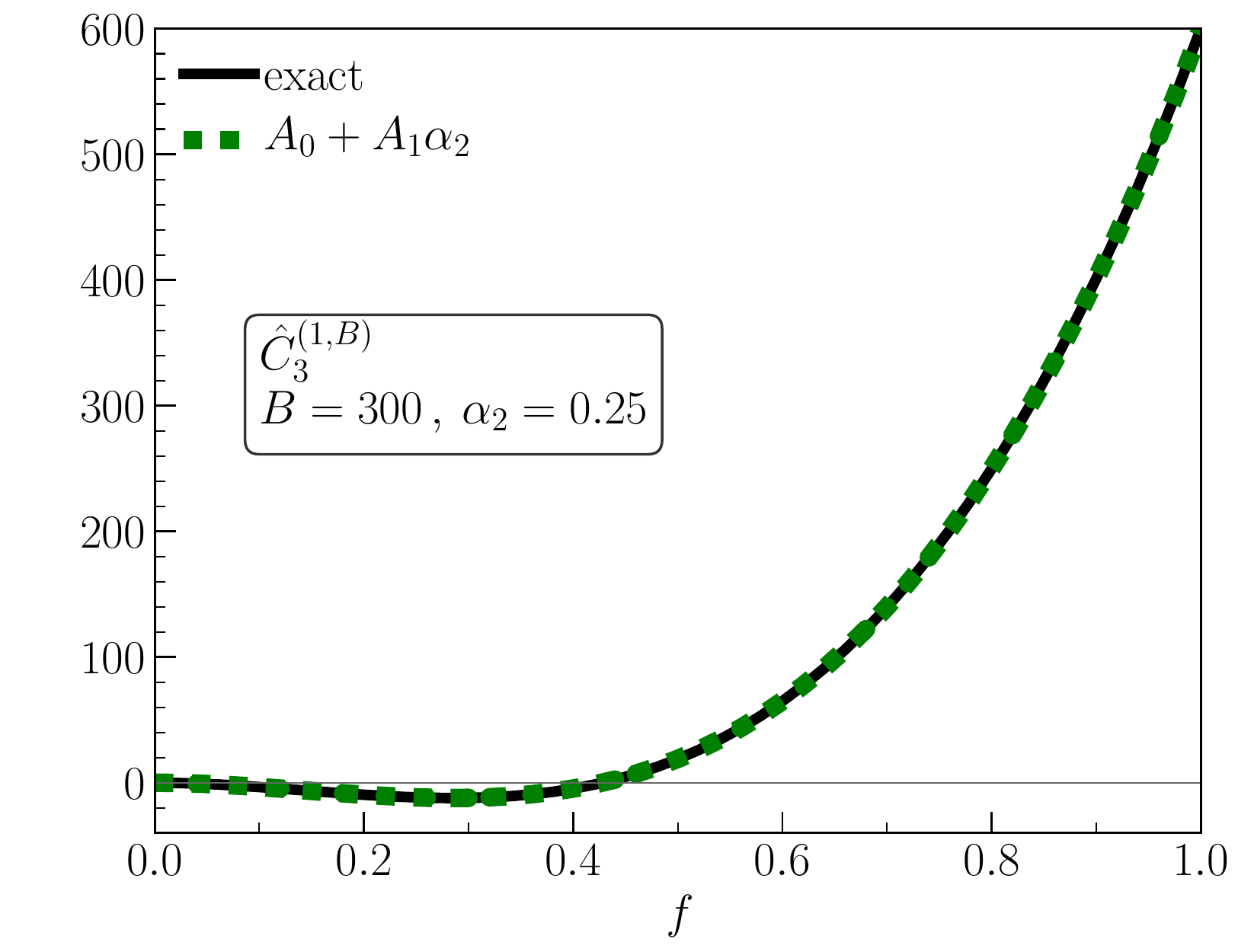}
	\includegraphics[width=0.49\textwidth]{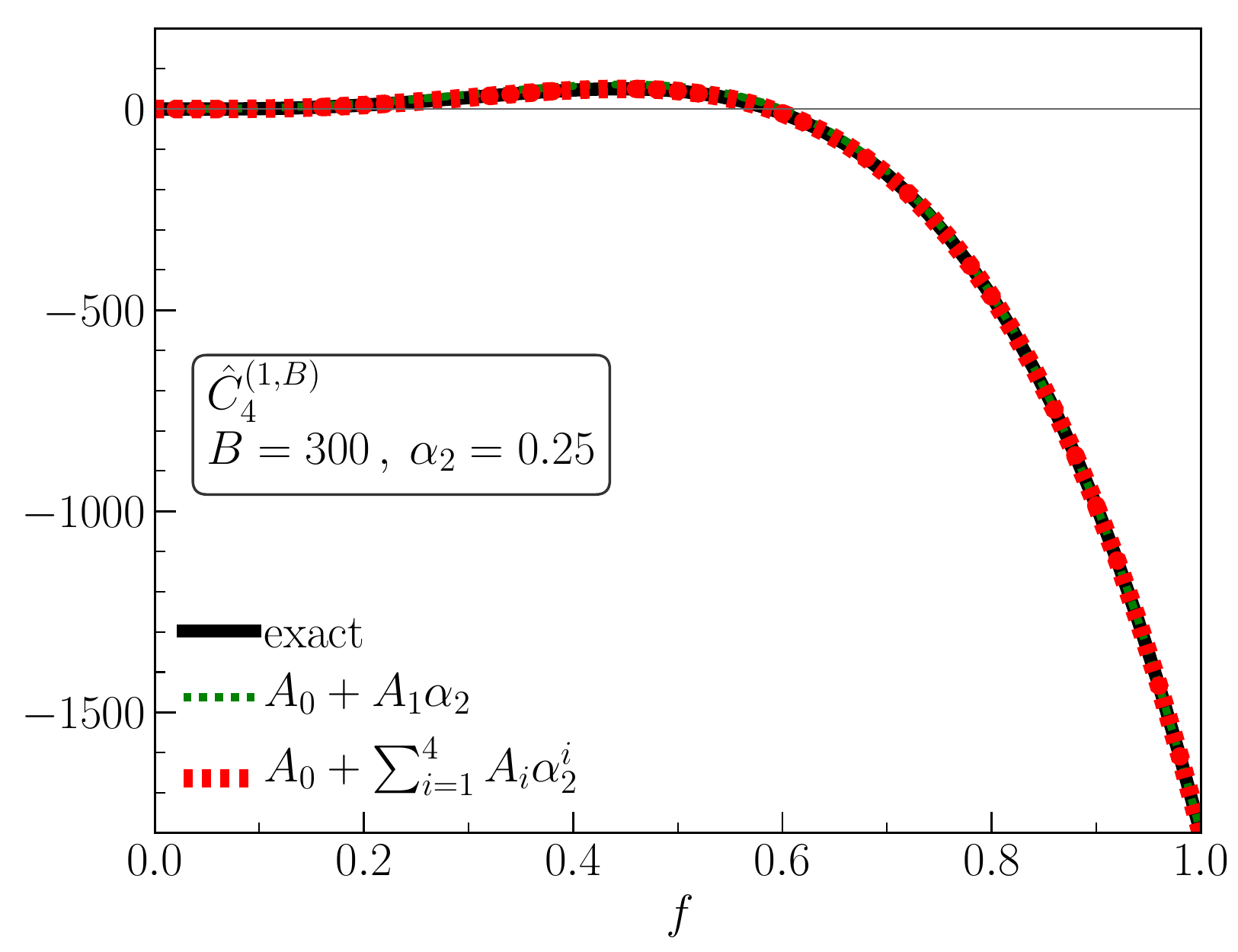}
	\includegraphics[width=0.49\textwidth]{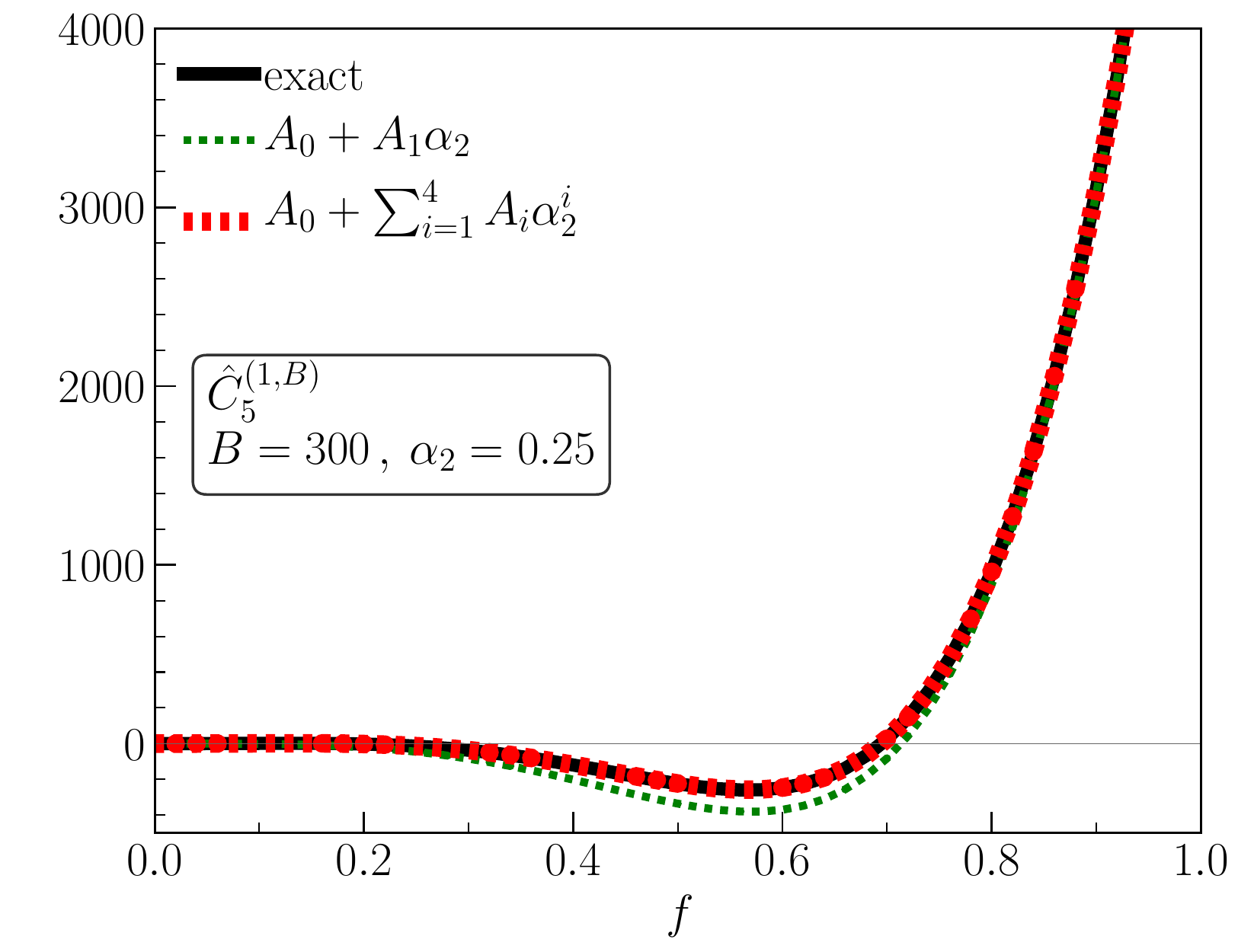}
	\includegraphics[width=0.49\textwidth]{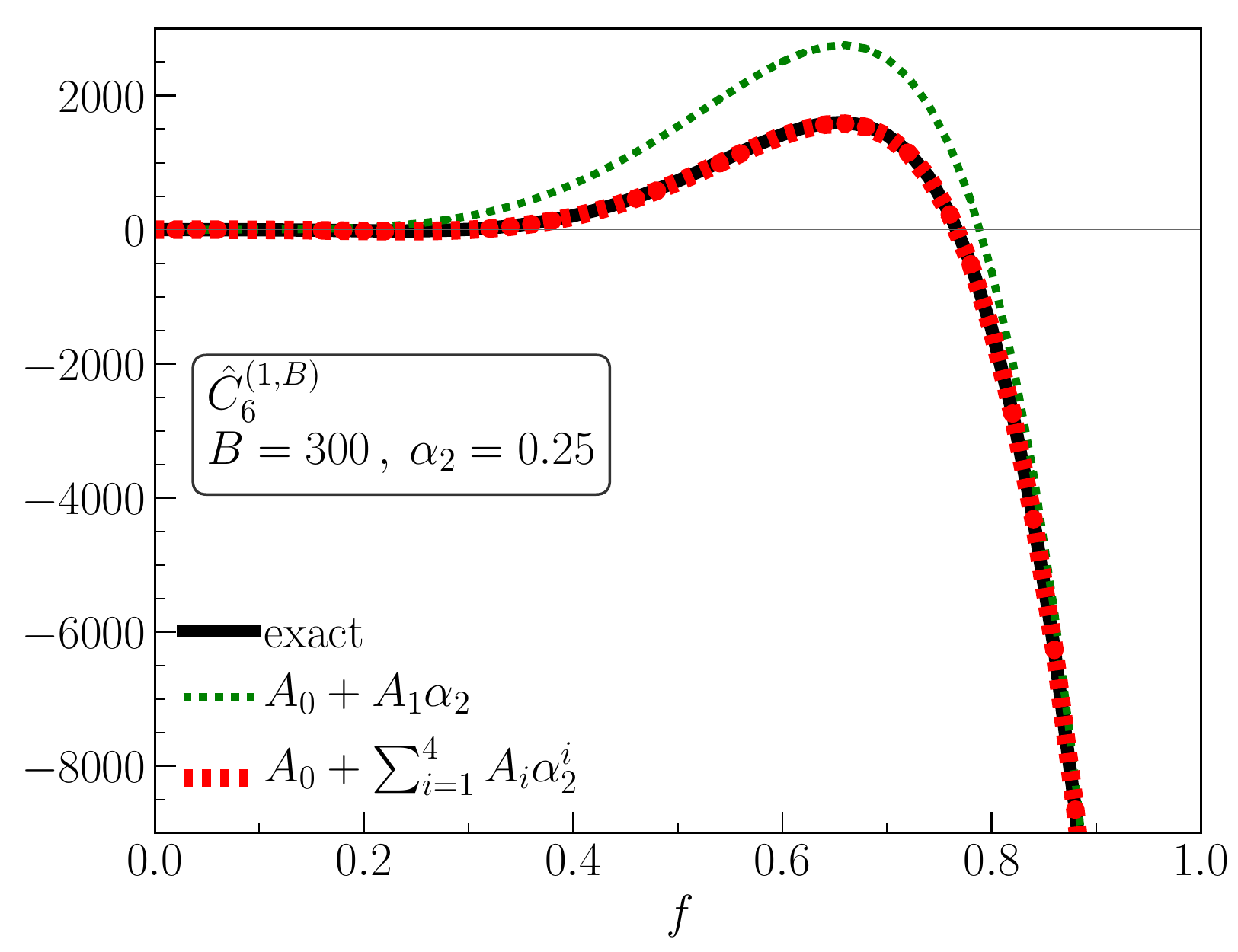}
\caption{Factorial cumulants $\hat{C}_n^{(1,B)}$ ($n=2,3,...,6$, assuming $B=300$, $\alpha_2=0.25$, and $\alpha_k = 0$ for $k \geq 3$) as a function of a fraction of particles in the first of the two subsystems, $f$. In each figure we show the ``exact'' line representing analytic results, Eqs. (\ref{eq:c1-bell})-(\ref{eq:c6-bell}), and approximate expansions from Eqs. (\ref{eq:c1})-(\ref{eq:c6}). For readability, we present only linear and quartic functions.} \label{fig:c-vs-f}
\end{center}
\end{figure}

It is also useful to plot these factorial cumulants as functions of $f$ for fixed $B$ and $\alpha_2$ (we choose $B=300$ and $\alpha_2=0.25$), see Fig. \ref{fig:c-vs-f}. For  $\hat{C}_2^{(1,B)}$, $\hat{C}_3^{(1,B)}$, and $\hat{C}_4^{(1,B)}$ already a linear expansion in $\alpha_2$ is in good agreement with the exact results. In the case of $\hat{C}_5^{(1,B)}$ and $\hat{C}_6^{(1,B)}$ significant deviations between the linear $\alpha_2$ expansion and the exact function are observed. Including higher order terms up to $\alpha_2^4$ is sufficient to reproduce the exact results. 

\section{Multiparticle correlations} \label{sec:multi-particle}

In this section we calculate 
the factorial cumulants taking into account the multiparticle short-range correlations. 
The generating function, Eq. (\ref{eq:g_b}), can be written as:

\begin{equation}\label{eq:g_b-for-many}
\begin{aligned}
G_{(1,B)}(z) = \ln\biggl[\frac{A}{B!} \biggl.\frac{d^B}{dx^B}& \biggl[ \exp\ml((xz-1)f \langle N \rangle + (x-1)\mfb\langle N \rangle \mr) \left(\sum_{m=0}^{\infty} \frac{V^m}{m!} \right) \biggr]\biggr|_{x=0}\biggr] \,,
\end{aligned}
\end{equation}
where 
\begin{equation}
V = \sum_{k=2}^{5} \left( \frac{(xz-1)^k}{k!}f\langle N \rangle \alpha_k + \frac{(x-1)^k}{k!}\mfb\langle N \rangle \alpha_k \right)\,.
\end{equation}
Here we assumed that $\alpha_k \neq 0$ for $k \leq 5$. 

Considering $\alpha_k$ to be small we can limit our expansion to a linear term in $V$:
\begin{equation}\label{eq:g_b-for-many-2}
G_{(1,B)}(z) \approx \ln\biggl[\frac{A}{B!} \biggl.\frac{d^B}{dx^B} \biggl[ \exp\ml((xz-1)f \langle N \rangle + (x-1)\mfb\langle N \rangle \mr) \left( 1 + V \right) \biggr]\biggr|_{x=0}\biggr] \,.
\end{equation}

By evaluating the derivatives using the general Leibnitz formula, we obtain:
\begin{equation} \label{eq:gb-linear}
G_{(1,B)}(z)\approx \ln(A)-\ln(B!)-\langle N \rangle + B \ln(\langle N \rangle Y_1) + \ln\biggl[1 + \langle N \rangle \sum_{k=2}^{5}A_{k} \alpha_k \biggr]\,, 
\end{equation}
where 
\begin{equation} \label{eq:gb-linear-series-terms}
A_{k} =\frac{1}{k!}  \sum_{j=0}^{k} (-1)^{k-j} \binom{k}{j} \frac{B!}{(B-j)!} \frac{X_j}{\langle N \rangle^j}  \,, 
\end{equation}
and $X_j$, $Y_1$ are defined below Eq. (\ref{eq:2-particle-gfun}). 

The factorial cumulants are given by (here $\langle N  \rangle = B$ and the calculated factorial cumulants are expanded in small $\alpha_k$)
\begin{fleqn} 
\begin{gather}
	\begin{aligned} \label{eq:many-c1}
	\hat{C}_1^{(1,B)} = fB  \,,
	\end{aligned}\\
	\begin{aligned}\label{eq:many-c2}
	\hat{C}_2^{(1,B)} &= -f^2B + \mfb f(B-1)\biggl[\;\alpha_2 -2\frac{1}{B} \; \alpha_3 -\frac{1}{2} \frac{B-6}{B^2}\;\alpha_4 + \frac{4}{3}\frac{B-3}{B^3}\;\alpha_5 \biggr] + \ldots  \,,
	\end{aligned}\\
	\begin{aligned}
	\hat{C}_3^{(1,B)} &= 2f^3B - \mfb f(B-1)\biggl[6f\;\alpha_2 +\frac{(2f-1)B + 2(1-8f)}{B} \; \alpha_3 
	 -3\frac{(3f-1)B+2(1-5f)}{B^2}\alpha_4 \\& \phantom{2f^3B - \mfb f\frac{B-1}{B^2}[ + + } \; +\frac{1}{2}\frac{(1-2f)B^2 + 2(22f-7)B + 24(1-4f)}{B^3}\alpha_5  \biggr] + \ldots  \,,
	\end{aligned}\\
	\begin{aligned}\label{eq:many-c4}
	\hat{C}_4^{(1,B)} &= -6f^4B + \mfb f(B-1)\biggl[36f^2\;\alpha_2 +12f\frac{(2f-1)B + 2(1-5f)}{B} \; \alpha_3 \\& \phantom{2f^3B - \mfb f\frac{B-1}{B^2}[} \; + \frac{B^2(3f^2-3f+1) -B(105f^2 - 51f + 5) + 6(45f^2 - 15f +1)}{B^2}\; \alpha_4 \\&
	 \phantom{2f^3B - \mfb f\frac{B-1}{B^2}} \; - 2\frac{B^2(12f^2-9f+2) -2B(69f^2 - 36f + 5) + 12(21f^2 - 9f +1)}{B^3}\; \alpha_5 \biggr] + \ldots  \,,
	\end{aligned}
\end{gather}
\end{fleqn}

Note that the $\alpha_2$ terms are in agreement with the linear part of Eqs. (\ref{eq:c1})-(\ref{eq:c4}). The higher order factorial cumulants can be also readily derived.

In the limit of large $B$ (and small $\alpha_k$) the factorial cumulants read\footnote{Here we present results up to $\hat{C}_6^{(1,B)}$ and thus we take $\alpha_k \neq 0$ for $k \leq 7$.}:
\begin{flalign}\label{eq:c1-many-b-large}
	\hat{C}_1^{(1,B)} &= fB  \,,&&
\end{flalign}
\begin{flalign}\label{eq:c2-many-b-large}
	\hat{C}_2^{(1,B)} &\approx fB\ml[ -f + \mfb \alpha_2 \mr]  \,,&&
\end{flalign}
\begin{flalign}\label{eq:c3-many-b-large}
	\hat{C}_3^{(1,B)} &\approx f B \ml[ 2f^2 -6 \mfb f \alpha_2 + \mfb (1-2f) \alpha_3 \mr]   \,,&&
\end{flalign}
\begin{flalign}\label{eq:c4-many-b-large}
	\hat{C}_4^{(1,B)} &\approx fB \; [ -3!f^3 +36\mfb  f^2 \alpha_2 -12\mfb  f(1-2f) \alpha_3 +\mfb (1-3\mfb f) \alpha_4 ] \,,&&
\end{flalign}
\begin{flalign}\label{eq:c5-many-b-large}
	\hat{C}_5^{(1,B)} \approx& f B [4! f^4 - 240 f^3 \mfb \alpha_2 + 120 f^2 \mfb (1 - 2 f) \alpha_3 - 20 f \mfb (1 - 3 f \mfb) \alpha_4 \\ \nonumber
		& \phantom{f B[} + \mfb (1 - 2 f) (1 - 2 f \mfb) \alpha_5 ]\,,&&
\end{flalign}
\begin{flalign}\label{eq:c6-many-b-large}
	\hat{C}_6^{(1,B)} \approx& f B [-5! f^5 + 1800f^4 \mfb \alpha_2 -1200f^3 \mfb (1-2f) \alpha_3 + 300f^2 \mfb (1-3f \mfb) \alpha_4 \\ \nonumber
	& \phantom{f B [} - 30f \mfb (1-2f)(1-2f \mfb) \alpha_5 + \mfb(1-5f \mfb(1-f \mfb)) \alpha_6 ] \,.&&
\end{flalign}

One can observe that for large $B$, $\hat{C}_n^{(1,B)}$ is not influenced by $\alpha_k$ with $k > n$. For example, in $\hat{C}_3^{(1,B)}$ only two- and three-particle short-range correlations, represented by $\alpha_2$ and $\alpha_3$, are significant and the higher-ordered ones are suppressed.  
In this calculation we assumed that $\alpha_k$ is small enough to include in Eq. (\ref{eq:g_b-for-many}) only the linear term in $V$. It was checked that our conclusion about the suppression of higher order $\alpha_k$ is also true if higher powers of $V$ are taken into account. To demonstrate this point, in Appendix \ref{appendix:appB} we present results with $(1 + V + V^2/2)$ instead of $(1+V)$ in Eq. (\ref{eq:g_b-for-many-2}).

\section{Agreement with Ref. \cite{Vovchenko:2020tsr}} \label{sec:confront}
It would be interesting to test our technique and reproduce the results presented in Ref. \cite{Vovchenko:2020tsr}.
We take \Crefrange{eq:c1-many-b-large}{eq:c4-many-b-large}, valid for large $B$, and known relations between cumulants and factorial cumulants, and calculate the cumulants in the first subsystem with short-range correlations and baryon number conservation. We obtain:
\begin{fleqn} 
\begin{gather}
	\begin{aligned} 
	\kappa_2^{(1,B)} \approx \mfb f B(\alpha_2 + 1)  \,,
	\end{aligned} \\
	\begin{aligned} 
	\kappa_3^{(1,B)} \approx \mfb f (1-2f)B(1 + 3\alpha_2 + \alpha_3)  \,,
	\end{aligned}\\
	\begin{aligned} 
	\kappa_4^{(1,B)} \approx \mfb f B[(1 + 7\alpha_2 + 6\alpha_3 + \alpha_4) - 3\mfb f (2 + 12\alpha_2 + 8\alpha_3 + \alpha_4)]  \,.
	\end{aligned}
\end{gather}
\end{fleqn}
The global factorial cumulants, that is in both subsystems from Fig. \ref{fig:box} combined, are given by $\hat{C}_n^{(G)} = B\alpha_n$. These factorial cumulants are defined before baryon number conservation is included (compare with Eq. (\ref{eq:corrs})).   
Using again the relations between cumulants and factorial cumulants we obtain
\begin{fleqn} 
\begin{gather}
	\begin{aligned} \label{eq:k2-1}
	\kappa_2^{(1,B)} \approx \mfb f \kappa_2^{(G)}  \,,
	\end{aligned} \\
	\begin{aligned} \label{eq:k3-1}
	\kappa_3^{(1,B)} \approx \mfb f (1-2f)\kappa_3^{(G)}  \,,
	\end{aligned}\\
	\begin{aligned} \label{eq:k4-1}
	\kappa_4^{(1,B)} \approx \mfb f [\kappa_4^{(G)} - 3\mfb f (\kappa_4^{(G)} + 2\kappa_3^{(G)} - \kappa_2^{(G)})]  \,,
	\end{aligned}
\end{gather}
\end{fleqn}
where $\kappa_n^{(G)}$ is the $n$-th global cumulant in the whole system originating from the short-range correlations but without the conservation of baryon number. $\kappa_n^{(1,B)}$ are the cumulants in one subsystem with all sources of correlations. 

Equations for $\kappa_2^{(1,B)}$ and $\kappa_3^{(1,B)}$ 
reproduce the relations obtained in Ref. \cite{Vovchenko:2020tsr}. We note here that in deriving \Crefrange{eq:k2-1}{eq:k4-1} we considered only terms linear in $\alpha_k$ (we started with generating function (\ref{eq:g_b-for-many-2})). We checked that taking higher $V$ terms in Eq. (\ref{eq:g_b-for-many}) is not changing $\kappa_2^{(1,B)}$ and $\kappa_3^{(1,B)}$ and thus this is the final result. However, the higher order $V$ terms change $\kappa_4^{(1,B)}$ and to reach agreement with Ref. \cite{Vovchenko:2020tsr} we take  
\begin{equation} \label{eq:gfun-multi-more-terms}
G_{(1,B)}(z) \approx \ln\biggl[\frac{A}{B!} \biggl.\frac{d^B}{dx^B} \biggl[ \exp\ml((xz-1)f \langle N \rangle + (x-1)\mfb\langle N \rangle \mr) \left(\sum_{m=0}^{M} \frac{V^m}{m!} \right) \biggr]\biggr|_{x=0}\biggr] \,
\end{equation} 
instead of Eq. (\ref{eq:g_b-for-many-2}). Next, we calculate $\kappa_4^{(1,B)}\kappa_2^{(1,B)}$ in the large $B$ limit taking $M=1,2,3,...$. We observed that for $M \ge 2$ the result is always given by
\begin{equation} 
	\kappa_4^{(1,B)} \kappa_2^{(1,B)} \approx \mfb^2 f^2 [\kappa_4^{(G)} \kappa_2^{(G)} - 3\mfb f ((\kappa_3^{(G)})^2 + \kappa_4^{(G)} \kappa_2^{(G)})] \,,
\end{equation}
and thus the formula for the fourth cumulant reads
\begin{flalign}\label{eq:k4-many-b-large}
	\kappa_4^{(1,B)} &\approx \mfb f \ml[ \kappa_4^{(G)} - 3\mfb f \ml(\kappa_4^{(G)} + \mf{(\kappa_3^{(G)})^2}{\kappa_2^{(G)}} \mr)  \mr]  \,,&&
\end{flalign}
which is in agreement with Ref. \cite{Vovchenko:2020tsr}. 

In a similar way we also calculated the large $B$ limit of $\kappa_5^{(1,B)}\kappa_2^{(1,B)}$ and $\kappa_6^{(1,B)}(\kappa_2^{(1,B)})^3$.\footnote{For the case of $\kappa_6^{(1,B)} (\kappa_2^{(1,B)})^3$ it was necessary to allow for $\alpha_6 \neq 0$.} We found that $\kappa_5^{(1,B)}\kappa_2^{(1,B)}$ is not changing for $M \ge 2$ and $\kappa_6^{(1,B)}(\kappa_2^{(1,B)})^3$ is not changing for $M \ge 4$. We obtain
\begin{flalign}\label{eq:k5-many-b-large}
	\kappa_5^{(1,B)} &\approx \mfb f (1-2f)\ml[(1-2\mfb f ) \kappa_5^{(G)} - 10\mfb f \mf{\kappa_3^{(G)}\kappa_4^{(G)}}{\kappa_2^{(G)}} \mr]  \,,&&
\end{flalign}
\begin{flalign}\label{eq:k6-many-b-large}
	\kappa_6^{(1,B)} &\approx \mfb f \{1-5\mfb f [1-\mfb f ]\}\kappa_6^{(G)} \\ \nonumber 
	&+ 5f^2 \mfb^2 \biggl\{ 3\mfb f  \ml( \mf{\kappa_3^{(G)}}{\kappa_2^{(G)}} \mr)^2 \mf{3\kappa_4^{(G)} \kappa_2^{(G)}-(\kappa_3^{(G)})^2}{\kappa_2^{(G)}} - 2(1-2f)^2 \mf{(\kappa_4^{(G)})^2}{\kappa_2^{(G)}} 
	- 3[1-3\mfb f ]\mf{\kappa_3^{(G)} \kappa_5^{(G)}}{\kappa_2^{(G)}} \biggr\} \,,&&
\end{flalign}
also in agreement with Ref. \cite{Vovchenko:2020tsr}.

\section{Comments and summary} \label{sec:comments}
In this paper we obtained the factorial cumulant generating function in one of the two subsystems (Eqs. (\ref{eq:g_b}) and (\ref{eq:g_b-bell})) assuming global baryon number conservation and short-range correlations. For simplicity, the case of one species of particles was discussed. Using this function, we calculated the factorial cumulants assuming two-particle short-range correlations (\Crefrange{eq:c1-bell}{eq:c6-bell} and \Crefrange{eq:c1}{eq:c6}). We showed how they depend on the correlation strength and acceptance and compared the approximated formulas with the exact ones (Figs. \ref{fig:c-vs-alpha} and \ref{fig:c-vs-f}). 

Next, we obtained expressions for the factorial cumulants assuming small multiparticle short-range correlations (\Crefrange{eq:many-c1}{eq:many-c4}), and we studied the limit of large baryon number $B$ (\Crefrange{eq:c1-many-b-large}{eq:c6-many-b-large}). It turns out that for the $n$-th factorial cumulant only short-range correlations of less or equal to $n$ particles are significant. Finally, we calculated cumulants and checked that for large $B$ they are in agreement with the results presented in Ref. \cite{Vovchenko:2020tsr}.

There are many ways to broaden our study. First, it would be interesting to calculate the next correction to the results from Ref. \cite{Vovchenko:2020tsr}, 
see, e.g., Eq. (\ref{eq:k4-many-b-large}). 
These results take the leading term in $B$ which is justified for large systems and in our approach it is possible to obtain higher order terms. Next, it would be interesting to expand our calculations to many species of particles. Finally, it is desired to investigate the convolution of baryon number conservation with other long-range correlations. This might be rather challenging.

\begin{acknowledgements}
We thank Volker Koch for useful discussions. This work was partially supported by the Ministry of Science and Higher Education, and by the National Science Centre, Grant No. 2018/30/Q/ST2/00101.
\end{acknowledgements}

\appendix 
\section{The Faà di Bruno's formula and the complete exponential Bell polynomials} \label{appendix:appBruno}
The Faà di Bruno's formula reads
\begin{equation} 
\mf{d^B}{dx^B} f(g(x)) = \sum_{i=1}^{B}f^{(i)} (g(x))\: {\rm{Bell}}_{B,i}(g'(x), g''(x), ..., g^{(B-i+1)}(x)) \,,
\end{equation}
where ${\rm{Bell}}_{B,i}(x_1, x_2, ..., x_{B-i+1})$ are the partial exponential Bell polynomials.

Applying this to Eq. (\ref{eq:g_b}) with
{
\begin{fleqn} 
\begin{gather}
	\begin{aligned}
	f(g(x)) = e^{g(x)} \,,
	\end{aligned}\\
	\begin{aligned}
	g(x) = \sum_{k=1}^{\infty} \frac{(xz-1)^k \hat{C}_k^{(1)} + (x-1)^k \hat{C}_k^{(2)}}{k!}\,,
	\end{aligned}
\end{gather}
\end{fleqn}
}
we obtain:
\begin{equation} 
\begin{aligned}
\frac{d^B}{dx^B} & \left. \left[\exp\left( \sum_{k=1}^{\infty} \frac{(xz-1)^k \hat{C}_k^{(1)} + (x-1)^k \hat{C}_k^{(2)}}{k!} \right) \right]  \right|_{x=0} \\
&= \sum_{i=1}^{B}\exp \left( \sum_{k=1}^{\infty} \frac{(xz-1)^k \hat{C}_k^{(1)} + (x-1)^k \hat{C}_k^{(2)}}{k!} \right)  \\  
& \phantom{\sum_{i=1}^{B}} \times  {\rm{Bell}}_{B,i}\biggl( \sum_{k=1}^{\infty} \frac{k z(xz-1)^{k-1} \hat{C}_k^{(1)} + k(x-1)^{k-1} \hat{C}_k^{(2)}}{k!}, \\
& \phantom{\sum_{i=1}^{B} \times {\rm{Bell}}_{B,i}\biggl(}    \sum_{k=2}^{\infty} \frac{k(k-1)z^2(xz-1)^{k-2} \hat{C}_k^{(1)} + k(k-1)(x-1)^{k-2} \hat{C}_k^{(2)}}{k!}, \\
& \phantom{\sum_{i=1}^{B} \times  {\rm{Bell}}_{B,i}\biggl(} \ldots, \\
&  \left. \sum_{k=B-i+1}^{\infty} \frac{k(k-1)\cdots(k-B+i)[z^{B-i+1}(xz-1)^{k-B+i-1} \hat{C}_k^{(1)} + (x-1)^{k-B+i-1} \hat{C}_k^{(2)}]} {k!} \biggr) \right|_{x=0}\,.
\end{aligned}
\end{equation}

After simplifications and evaluation at $x=0$ we obtain:
\begin{equation} 
\begin{aligned}
\frac{d^B}{dx^B} & \left. \left[\exp\left( \sum_{k=1}^{\infty} \frac{(xz-1)^k \hat{C}_k^{(1)} + (x-1)^k \hat{C}_k^{(2)}}{k!} \right) \right]  \right|_{x=0} = \\ 
	& C \sum_{i=1}^{B} {\rm{Bell}}_{B,i} \biggl(  \sum_{k=0}^{\infty}\mf{(-1)^k}{k!} \ml[\hat{C}_{k+1}^{(1)}z + \hat{C}_{k+1}^{(2)}\mr], \\
	& \phantom{C \sum_{i=1}^{B} {\rm{Bell}}_{B,i} \biggl(} \sum_{k=0}^{\infty}\mf{(-1)^k}{k!} \ml[\hat{C}_{k+2}^{(1)}z^2 + \hat{C}_{k+2}^{(2)}\mr],  \\
	& \phantom{C \sum_{i=1}^{B} {\rm{Bell}}_{B,i} \biggl(} \ldots, \\
	& \phantom{C \sum_{i=1}^{B} {\rm{Bell}}_{B,i} \biggl(} \sum_{k=0}^{\infty}\mf{(-1)^k}{k!} \ml[\hat{C}_{k+B-i+1}^{(1)} z^{B-i+1} + \hat{C}_{k+B-i+1}^{(2)}\mr] \biggr) \biggr] \,.
\end{aligned}
\end{equation}

Here $C = \exp \ml[\sum_{k=1}^{\infty} \mf{(-1)^k}{k!} \ml( \hat{C}_k^{(1)} + \hat{C}_k^{(2)} \mr) \mr] $ is a constant w.r.t. $z$ which is unimportant for calculations of the factorial cumulants. Then by applying the relation between the complete and partial exponential Bell polynomials, see Eq. (\ref{eq:complete-bell}), and denoting $A' = A C$, we obtain Eq. (\ref{eq:g_b-bell}).

\section{Multi-particle factorial cumulants from $(1 + V + \mf{1}{2} V^2)$}
\label{appendix:appB}
Instead of Eq. (\ref{eq:g_b-for-many-2}), we use the following factorial cumulant generating function:
\begin{equation}\label{eq:g_b-for-many-2-quadr}
G_{(1,B)}(z) \approx \ln\biggl[\frac{A}{B!} \biggl.\frac{d^B}{dx^B} \biggl[ \exp\ml[(xz-1)f  B  + (x-1)\mfb B  \mr] \left( 1 + V + \mf{1}{2}V^2 \right) \biggr]\biggr|_{x=0}\biggr]\,,
\end{equation}
where $\alpha_k \neq 0$ for $k \leq 7$ and
\begin{equation}
V = \sum_{k=2}^{7} \left( \frac{(xz-1)^k}{k!}f B \alpha_k + \frac{(x-1)^k}{k!}\mfb B \alpha_k \right)\,.
\end{equation}

The factorial cumulants in the limit of large $B$ read:
\begin{flalign}\label{eq:c1-many-b-large-quadr}
	\hat{C}_1^{(1,B)} &= fB  \,,&&
\end{flalign}
\begin{flalign}\label{eq:c2-many-b-large-quadr}
	\hat{C}_2^{(1,B)} &\approx fB\ml[ -f + \mfb \alpha_2 \mr]  \,,&&
\end{flalign}
\begin{flalign}\label{eq:c3-many-b-large-quadr}
	\hat{C}_3^{(1,B)} &\approx f B \ml[ 2f^2 -6 \mfb f \alpha_2 + \mfb (1-2f) \alpha_3 \mr]   \,,&&
\end{flalign}
\begin{flalign}\label{eq:c4-many-b-large-quadr}
	\hat{C}_4^{(1,B)} &\approx  fB [ -3!f^3 +36\mfb  f^2 \alpha_2 -12\mfb  f(1-2f) \alpha_3 +\mfb (1-3\mfb f) \alpha_4 
	\\ \nonumber 
	&\phantom{= f B[} -12 \mfb^2 f \alpha_2^2 + -3\mfb^2 f \alpha_3^2  -12\mfb^2 f \alpha_2 \alpha_3 ]  \,.&&
\end{flalign}
\begin{flalign}\label{eq:c5-many-b-large-quadr}
	\hat{C}_5^{(1,B)} \approx& f B [4! f^4 - 240 f^3 \mfb \alpha_2 + 120 f^2 \mfb (1 - 2 f) \alpha_3 - 20 f \mfb (1 - 3 f \mfb) \alpha_4 \\ \nonumber
	& \phantom{f B[} + \mfb (1 - 2 f) (1 - 2 f \mfb) \alpha_5 \\ \nonumber
    & \phantom{f B[} + 240 f^2 \mfb^2  \alpha_2^2 - 30 f \mfb^2 (1 - 4 f) \alpha_3^2 - 60 f \mfb^2 (1 - 6 f) \alpha_2 \alpha_3
    - 20 f \mfb^2 (1 - 2 f) \alpha_2 \alpha_4  \\ \nonumber
    & \phantom{f B [}- 10 f \mfb^2 (1 - 2 f) \alpha_3 \alpha_4 ]   \,,&&
\end{flalign}
\begin{flalign}\label{eq:c6-many-b-large-quadr}
	\hat{C}_6^{(1,B)} \approx& f B [-5! f^5 + 1800f^4 \mfb \alpha_2 -1200f^3 \mfb (1-2f) \alpha_3 + 300f^2 \mfb (1-3f \mfb) \alpha_4 \\ \nonumber
	& \phantom{f B [} - 30f \mfb (1-2f)(1-2f \mfb) \alpha_5 + \mfb(1-5f \mfb(1-f \mfb)) \alpha_6 \\ \nonumber
	& \phantom{f B [} - 3600f^3 \mfb^2 \alpha_2^2 -90f \mfb^2 (1-14f+34f^2) \alpha_3^2 - 10f \mfb^2 (1-2f)^2 \alpha_4^2 \\ \nonumber
	& \phantom{f B [} + 1800f^2 \mfb^2(1-4f)\alpha_2 \alpha_3 -120f \mfb^2(1-8f+13f^2)\alpha_2 \alpha_4 \\ \nonumber
	& \phantom{f B [} - 60f \mfb^2(2-12f+17f^2)\alpha_3 \alpha_4 - 30f \mfb^2 (1-3f \mfb) \alpha_2 \alpha_5 - 15f \mfb^2 (1-3f \mfb)\alpha_3 \alpha_5 ]   \,.&&
\end{flalign}

We note that the $V^{2}$ term is not affecting $\hat{C}_1^{(1,B)}$, $\hat{C}_2^{(1,B)}$, and $\hat{C}_3^{(1,B)}$ whereas we have higher order terms in $\hat{C}_4^{(1,B)}$, $\hat{C}_5^{(1,B)}$, and $\hat{C}_6^{(1,B)}$. Importantly, the observation that the $\alpha_k$ terms in $\hat{C}_n^{(1,B)}$ where $k > n$ are suppressed, is confirmed here. We also checked up to $(1+V+ \mf{1}{2}V^2 + \mf{1}{3!}V^3 + \mf{1}{4!}V^4)$ that this conclusion remains true.

\bibliography{v14.bbl}

\end{document}